\documentclass[a4paper,11pt]{article}
\usepackage{pos}
\usepackage{amsmath}
\usepackage{graphicx}
\usepackage{xcolor}
\usepackage{indentfirst}
\usepackage{amsfonts}
\usepackage{bm}
\usepackage{physics}

\title{Spectral sum from Euclidean lattice correlators and determination of renormalization constants}
\ShortTitle{Spectral sum and renormalization constants}

\author*[a,b,c]{Tsutomu Ishikawa}
\author[a,b]{Shoji Hashimoto}
\author[a,b]{Takashi Kaneko}

\affiliation[a]{High Energy Accelerator Research Organization (KEK), Tsukuba 305-0801, Japan}

\affiliation[b]{The Graduate University for Advanced Studies, SOKENDAI, Tsukuba 305-0801, Japan}

\affiliation[c]{RIKEN Center for Computational Science, Kobe 650-0047, Japan}

\emailAdd{tsutomu.ishikawa@a.riken.jp}
\abstract{
We propose a new method to renormalize lattice operators.
The method is based on the technique to compute the spectral sum appearing in the Shifman-Vainshtein-Zakharov QCD sum rule from lattice correlators.
The application of this technique to the light quark system is useful for operator renormalization as well as for the test of perturbative QCD and OPE.
We determine the renormalization constant of the vector current and discuss extensions to other current operators.}

\FullConference{%
 The 38th International Symposium on Lattice Field Theory, LATTICE2021
  26th-30th July, 2021
  Zoom/Gather@Massachusetts Institute of Technology
}

\begin{document}

\begin{flushright}
    KEK-CP-0388
\end{flushright} 

\maketitle

\section{Introduction}

Lattice QCD has been used to study hadronic decays and transition processes nonperturbatively. Measuring two- and three-point correlation functions, one determines parameters such as decay constants and form factors from hadronic matrix elements.
These parameters are important inputs for the phenomenological study of the Standard Model.

Renormalization is necessary in the calculation of hadronic matrix elements.
These matrix elements from lattice QCD do not correspond to the values in the renormalization scheme of the continuum theory (usually $\overline{\text{MS}}$).
Even if an operator involved in these matrix elements has no anomalous dimensions, such as locally constructed vector currents $\bar{q}\gamma_\mu q$, the renormalization has to be done since the current does not conserve due to lattice artifacts.
These physical quantities computed on the lattice can be used in the continuum theory after the  renormalization.

We can perform renormalization through matching.
The basic strategy is as follows.
We measure some physical quantity containing the operator in lattice QCD.
We calculate the same quantity in the $\overline{\text{MS}}$ scheme perturbatively. Then we determine the renormalization constant by requiring them to be equal.
Notice that it is also possible to match via an intermediate scheme such as the RI/MOM scheme \cite{Martinelli:1994ty}.
However, we focus on the direct renormalization method in this work.
The quantity used for the matching should be easy to control the discretization error in lattice QCD.
At the same time, the typical energy scale of the quantity has to be sufficiently large to use the perturbative expansion and operator product expansion (OPE).
If we find the quantity which satisfies these requirements, the renormalization constant will be less affected by the systematic error.

As a first step, we consider current-current correlators in momentum space as a quantity for the matching.
These correlators are defined as
\begin{align}
\label{eq:FT}
    \Pi_{\Gamma}(q^2) =i \int d^4x\, e^{i q x}\langle J_{\Gamma}(x) J_{\Gamma}(0) \rangle,
\end{align}
where $J_{\Gamma}$ is a bilinear operator such as $\bar{q} \gamma_\mu q$, $\bar{q} \gamma_5 q$, and $\bar{q} \sigma_{\mu\nu} q$.
In the case of the vector current, we can write this correlator as $\Pi_{V}(q^2)=(q_\mu q_\nu - q^2g_{\mu \nu})\Pi(q^2)$. The function $\Pi(q^2)$ is the so-called hadronic vacuum polarization (HVP) function.
In the deep Euclidean region $Q^2=-q^2\gg 0$, the perturbative expansion and OPE are applicable.
The perturbative series in the massless limit has been calculated up to $O(\alpha_s^3)$ for the vector current \cite{Chetyrkin:2010dx}.
Lattice calculation for the quantity is straightforward.
The convergences of the OPE is, however, problematic, as discussed in the literature \cite{Hudspith:2018bpz}.
Namely, there is a severe window problem $\Lambda_{\text{QCD}}^2\ll Q^2 \ll 1/a^2$. Therefore, the correlator in momentum space $\Pi_{\Gamma}(q^2)$ itself would not be suitable for the renormalization.

In the present work, we propose a new method to renormalize lattice operators using the Borel transformation, which is often utilized in the QCD sum rule. The method is based on the technique to compute the weighted spectrum in the numerical lattice computation \cite{Bailas:2020qmv}. We compute the renormalization constant and compare it with the result from another nonperturbative method.

\section{Borel transform of HVP}

In the following, we limit ourselves to the renormalization of the vector current for simplicity. The extension to other operators will be straightforward.
Some discussions of other channels are in Section \ref{sc:summary}.

To improve the convergence of OPE, we follow the Shifman-Vainshtein-Zakharov (SVZ) QCD sum rule \cite{Shifman:1978bx,Shifman:1978by}.
We use a weighted spectrum as employed in the sum rule studies:
\begin{align}
\label{eq:Borel_sum}
    \tilde{\Pi}(M^2)&\equiv\frac{1}{M^2}\int_0^{\infty}ds\, \rho(s)e^{-s/M^2},\\
    \rho(s) &= \Im \Pi(s+i\epsilon),
\end{align}
where $\rho(s)$ is the spectral function, which is understood as the density of states of a given energy.
The weighted spectrum can be derived from the HVP through the Borel transformation
\begin{align}
    \mathcal{B}_M=\lim_{\substack{n,Q^2\rightarrow \infty \\ Q^2/n=M^2} }\frac{(Q^2)^{n}}{(n-1)!}\left(-\frac{\partial}{\partial Q^2}\right)^n.
\end{align}
Applying this transformation to the dispersion relation of HVP, one obtains
\begin{align}
    \mathcal{B}_M[\Pi(-Q^2)]=  \int_0^\infty ds\, \mathcal{B}_M\bqty{\frac{1}{s+Q^2}}\rho(s)=\frac{1}{M^2}\int_0^{\infty}ds\, \rho(s)e^{-s/M^2}.
\end{align}
The typical scale is the Borel mass $M^2$ in the exponential factor.
The perturbative series can also be transformed from two-point functions. For the vector and scalar operators, the corrections at $O(\alpha_s^4)$ are found in \cite{Chetyrkin:2010dx}. For the tensor currents, the corrections at $O(\alpha_s^3)$ are also in this reference.
An important property of the Borel transformation is the factorial suppression of the power corrections:
\begin{align}
    \mathcal{B}_M\bqty{\frac{1}{Q^{2n}}}=\frac{1}{(n-1)!}\frac{1}{M^{2n}}.
\end{align}
Therefore, the uncertainty due to higher-dimensional power corrections are substantially reduced.

\section{Method}

We will briefly explain the point of our method to compute the Borel transform of HVP $\tilde{\Pi}(M^2)$ from lattice QCD.
More details are in \cite{Ishikawa:2021txe}; the method yields reasonable agreement with OPE results in the $s\bar{s}$ system. 
Other applications based on the transfer matrix expansion can be found in \cite{Gambino:2020crt,Fukaya:2020wpp}.

The lattice computation of the Borel transform is nontrivial since its definition requires the spectral function.
We define a correlator with time separation $t$ as
\begin{align}
    C(t)=\sum_{\vb{x}}\expval{J_{z}(t,\vb{x})J_{z}(0,\vb{0})}{0},
\end{align}
which is projected to zero spatial momentum.
The correlator is also a spectral sum since we can express it as an integral \cite{Bernecker:2011gh}:
\begin{align}
\label{eq:spct_rep_C(t)}
    C(t)=\int_0^\infty d\omega\, \omega^2 \rho(\omega^2) e^{-\omega t},
\end{align}
where $\omega^2=s$ and $t>0$. 
The estimate of the spectral function $\rho(\omega^2)$ from \eqref{eq:spct_rep_C(t)} is an ill-posed problem.
One can extract only, at most, a few low-lying spectra from the correlators.
Instead, we compute the spectral sum \eqref{eq:Borel_sum} by the transfer matrix expansion without estimation of the spectral function itself. 

The transfer matrix $e^{-\hat{H}a}$ on the lattice plays the role of the discrete time evolution operator.
We formally write the correlator at a time separation $t=n a$ as a matrix element
\begin{align}
\label{eq:trsf_C(t)}
    C(t)&=\sum_{\vb{x}}\expval{J(0,\vb{x})\pqty{e^{-\hat{H}a}}^n J(0,\vb{0})}{0}.
\end{align}
According to \eqref{eq:spct_rep_C(t)}, we can express the spectral function by a delta function:
\begin{align}
\label{eq:spct_delta}
    \omega^2 \rho(\omega^2)&=\sum_{\vb{x}}  \expval{J(0,\vb{x})\delta(\hat{H}-\omega) J(0,\vb{0})}{0}.
\end{align}
The above two equations \eqref{eq:trsf_C(t)} and \eqref{eq:spct_delta} provide a clue to the computation of $\tilde{\Pi}(M^2)$.
Expanding the weight function,
\begin{align}
\label{eq:weight}
   S(M,\omega)\equiv  \frac{2C(2t_0)e^{2t_0\omega}}{M^2\omega}e^{-\omega^2/M^2}\tanh{(\omega/\omega_0)},
\end{align}
that correctly reproduces the integral \eqref{eq:Borel_sum}, we can relate the weighted spectral sum to the correlator
\begin{align}
    \frac{1}{M^2}\int_0^{\infty}ds\, \rho(s)e^{-s/M^2} &\simeq \int_0^\infty d\omega\, \sum_n a_n(M^2) \pqty{e^{-\omega a}}^n \omega^2\rho(\omega^2)\\
    &= \sum_n a_n(M^2) C(t),
\end{align}
where $a_n(M^2)$ is the coefficient of the expansion.

In actual computation, we use (shifted) Chebyshev expansion following \cite{Bailas:2020qmv}. The basis of this expansion is an orthogonal polynomial $T_j^*(x)$, whose absolute value is bounded by 1 in the range $0\leq x \leq 1$. Note that the expansion does not require $x=e^{-\omega a}$ to be small.
We obtain the Borel transform through the following equation with Chebyshev matrix elements $\expval*{T^*_j(e^{-\hat{H} a})}$:
\begin{align}
\label{eq:Borel_Cheb}
    \Pi(M^2)\simeq \frac{c_0^*(M^2)}{2}+\sum_{j=1}^{N_t}c_j^*(M^2)\expval*{T^*_j(e^{-\hat{H} a})},
\end{align}
where the coefficient $c_j^*$ is determined from the weight function by an integral, whose full expression can be found in \cite{Ishikawa:2021txe}. The contact term at $t=0$ and the divergence of the integral at $\omega=0$ are taken into account in \eqref{eq:weight}. 
The shifted Chebyshev polynomials is recursively constructed by $T_{j+1}^*(x)=(4x-2)T_j^*(x)-T_{j-1}^*(x)$ with $T_0^*(x)=1$ and  $T_1^*(x)=2x-1$.
After properly normalizing the correlator $\bar{C}(t)\equiv C(t)/C(0)$ so that $\bar{C}(0)=1$,
we can determine the Chebyshev matrix elements from lattice simulations as
\begin{align}
\label{eq:CME}
    \expval*{T^*_1(e^{-\hat{H} a})}=2\bar{C}(1)-1,\,\expval*{T^*_2(e^{-\hat{H} a})}=8\bar{C}(2)-8\bar{C}(1)+1,\,\ldots,
\end{align}
where we replace $x^n$ in $T_j^*(x)$ by $\bar{C}(n=t/a)$.
In practice, we determine the Chebyshev matrix element from a fit of \eqref{eq:CME}.
Note that the matrix elements are restricted by $|\expval*{T^*_j(e^{-\hat{H} a})}|\leq 1$. The restriction forbids the significant error due to large cancellation among the correlators at different time slices.
Combining $c_j^*(M^2)$ and $\expval*{T^*_j(e^{-\hat{H} a})}$, We compute the Borel transform $\tilde{\Pi}(M^2)$.

We culculate the correlator of the current $J_\mu=\bar{u}\gamma_\mu d$ at three lattice spacing.
We use ensembles with $N_f=2+1$ dynamical M\"obius domain-wall fermions, which are generated by the JLQCD collaboration.
Table \ref{tb:ensemble} shows the parameters of the ensembles, where light and strange sea quark masses are denoted by $m_{ud}$ and $m_s$, respectively.
We also use these degenerate masses $m_{ud}$ as valence quark masses.
The correlators are measured $N_{\text{meas}}$ times at each lattice spacing.
Some details of the configurations can be found in \cite{Nakayama:2016atf}.

\begin{table}[t]
 %\vspace{10pt}
 \centering
 \begin{tabular}{ccc|cccc}
   $\beta$ &   $ a^{-1}\ [\mathrm{GeV}]$  & $L^3\times T (\times L_5 )$ &$N_{\text{meas}}$ &$am_{ud}$&$am_s$\\ \hline
   4.17& 2.453(4)&$32^3 \times 64 (\times 12) $& 800& 0.0070&0.0400\\
   4.35& 3.610(9)&$48^3 \times 96 (\times 8) $& 600& 0.0042&0.0250\\
   4.47& 4.496(9)&$64^3 \times 128 (\times 8) $& 400& 0.0030&0.0150\\
\end{tabular}
\caption{Ensembles used in the present work.}
 \label{tb:ensemble}
\end{table}

As a test of our renormalization procedure, we compute the renormalization constant of the vector current. 
We impose the matching (renormalization) condition on the Borel transform of HVP, $\tilde{\Pi}^{\text{lat}}(a^2;M^2)$, computed at lattice spacing $a$:
\begin{align}
\label{eq:RG_cond}
    \tilde{\Pi}^{\overline{\text{MS}}}(\mu^2;M^2)&=\pqty{Z_{V}^{\overline{\text{MS}}/\text{lat}}(a^2)}^2\tilde{\Pi}^{\text{lat}}(a^2;M^2),
\end{align}
where $Z_V(a^2)$ is the renormalization constant for the vector current operator.
The l.h.s. of \eqref{eq:RG_cond}, $\tilde{\Pi}^{\overline{\text{MS}}}(\mu^2;M^2)$, denotes the counterpart in the perturbative QCD at the renormalization point $\mu$.
The perturbative expansion is known to $O(\alpha_s^4)$ in the massless limit.
For the vector current, there is no anomalous dimension, and the renormalization constant is independent of the renormalization scale up to the truncation error.
Since we obtain the renormalization condition \eqref{eq:RG_cond} at several $M^2$,
we obtain the optimal solution $Z_{V}^{\overline{\text{MS}}/\text{lat}}(a^2)$ as discussed in the next section.

\section{Result}
\begin{figure}[t]
\centering
  \includegraphics[width=12cm]{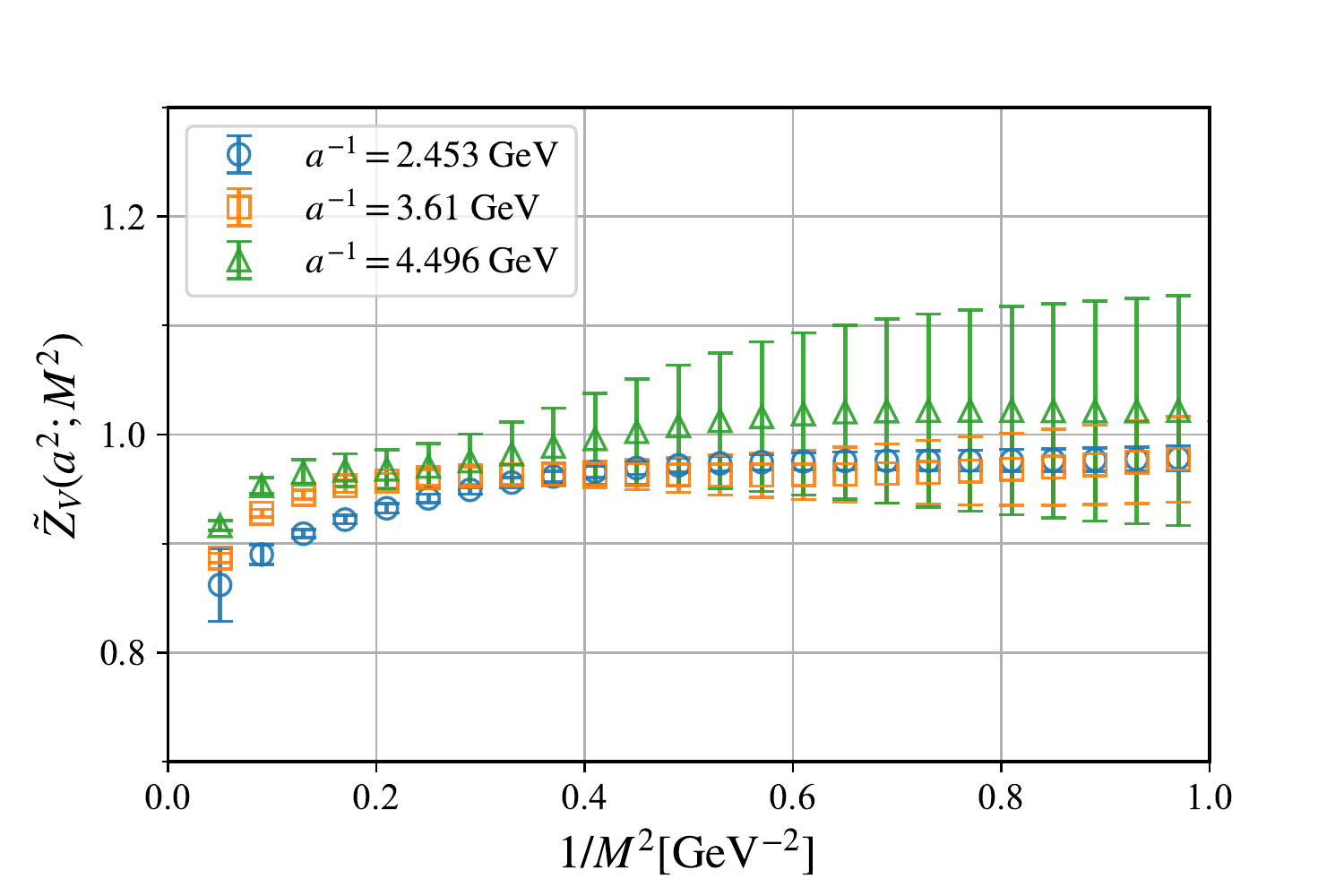}
 \small
 \caption{ $\tilde{Z}_V(a^2;M^2)$ at all lattice spacing. }\label{fig:M2_dep_latt}
\end{figure}

We show our preliminary results.
We set $N_t=18$ for the Borel transform in \eqref{eq:Borel_Cheb}. The perturbative expansion $\tilde{\Pi}^{\overline{\text{MS}}}(\mu^2;M^2)$ is computed in the massless limit, where we set $\mu=2$~GeV. 
Solving \eqref{eq:RG_cond} for $Z_{V}^{\overline{\text{MS}}/\text{lat}}$,
we can express the solution in the following form:
\begin{align}
\label{eq:ratio}
    \tilde{Z}_{V}(a^2;M^2)
    &\equiv
    \sqrt{\frac{\tilde{\Pi}^{\overline{\text{MS}}}(\mu^2;M^2)}{\tilde{\Pi}^{\text{lat}}(a^2;M^2)}}\\
    \label{eq:ansatz}
    &=Z_{V}^{\overline{\text{MS}}/\text{lat}}(a^2)+C_{-2}(Ma)^2+C_{4}/M^4.
\end{align}
The discretization effect is incorporated into this function as $C_{-2}(Ma)^2$. The term $C_{4}/M^4$ corresponds to the nonperturbative corrections due to the dimension-four operators.
Figure \ref{fig:M2_dep_latt} shows the $M^2$ dependence of $\tilde{Z}_{V}(a^2;M^2)$ at each lattice spacing.
In the short-distance region, namely at small $1/M^2$, the ratio  $\tilde{Z}_{V}(a^2;M^2)$ is suppressed due to the discretization effect. 
We determine the renormalization constant by a fit using the ansatz \eqref{eq:ansatz} in the range $1/M^2=0.25\text{--}0.69\,\text{GeV}^{-2}$. 
Then we obtain $Z_{V}^{\overline{\text{MS}}/\text{lat}}(a^2)=0.9804(43), 0.9806(27), 0.9789(23)$ at $a^{-1}=2.453, 3.610, 4.496$~GeV, respectively, where the parentheses denote the statistical errors only.

\begin{figure}[t]
\centering
  \includegraphics[width=12cm]{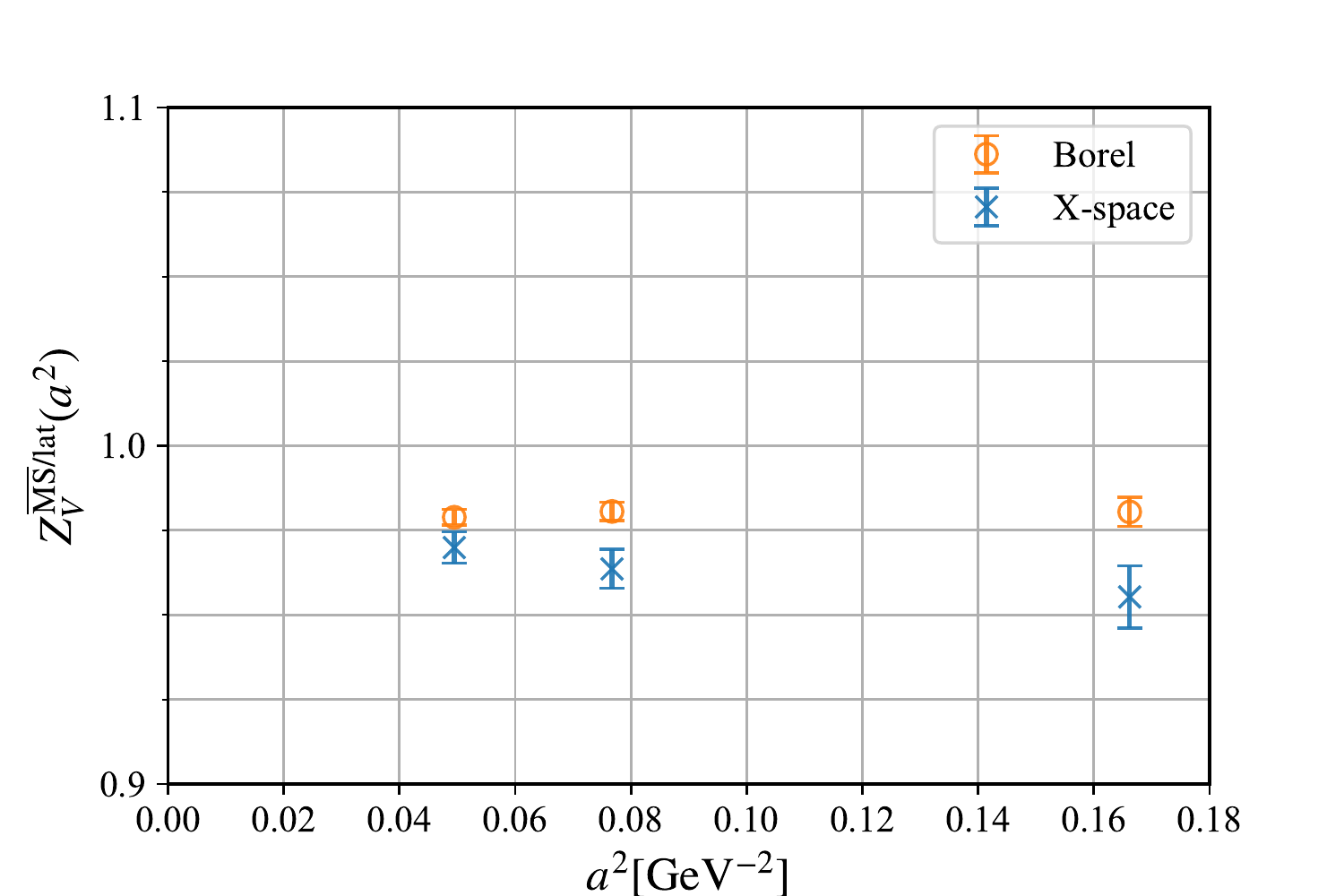}
 \small
 \caption{ Comparison of $Z_V^{\overline{\text{MS}}/\text{lat}}(a^2)$ at three lattice spacing. }\label{fig:a2_dep_ZV}
\end{figure}

We compare our result with another method in Fig. \ref{fig:a2_dep_ZV}. We take the renormalization constant from the X-space correlator as a reference \cite{Tomii:2016xiv}.
The horizontal axis in this figure is the squared lattice spacing.
The circle and cross symbols denote our result and the X-space method, respectively.
Note that the errors of our result does not include  systematic errors, such as truncation errors and valence quark mass dependence.
These renormalization constants may contain different discretization errors, and the maximal deviation is $\sim 2.6$\% on the coarse lattice.
The deviation becomes insignificant towards the continuum limit.

We are trying to extend our method to other operators, such as the (pseudo-)scalar density and tensor operators. Since those have finite anomalous dimensions, we have to give the correct scale dependence, unlike the vector current.
We compute the corresponding ratio such as \eqref{eq:ratio} but for tensor operators on the coarse lattice. 
We check that a ratio $\tilde{Z}_T(\mu^2,a^2;M^2)/\tilde{Z}_T(4~\text{GeV}^2,a^2;M^2)$ is consistent with the running at one-loop level. The computation is still ongoing.

The modification of the weight kernel is necessary for the pseudo-scalar density $\bar{u} \gamma_5 d$, because
the Borel transform is largely affected by $\pi$ meson since the exponential kernel does not suppress $\pi$ meson spectrum $\sim \delta(s-m_\pi^2)$. Accordingly, the Borel transform in this channel is not well described by perturbative QCD even at $M= 2$~GeV.
This would be improved by a replacement of the kernel $ e^{-s/M^2}\to se^{-s/M^2}$. The counterpart of this modified spectral sum in perturbative QCD may be derived by some mathematical manipulations, such as the derivative of the Borel transform:
\begin{align}
    \pdv{(1/M^2)}\mathcal{B}_M\bqty{\log^n \pqty{\frac{\mu^2}{Q^2}}}.
\end{align}

\section{Summary}
\label{sc:summary}
We propose a renormalization method based on the Borel transform following SVZ. The renormalization constant can be computed through the two-point correlation functions, which is fairly standard to compute in lattice calculation.
The perturbative expansion of this quantity is available.
The scale parameter $M^2$ is continuous and easily adjustable using the Chebyshev expansion.
The errors are due to the truncation of perturbative expansion, the finite mass correction, and the choice of the fit range as well as the Chebyshev expansion.
The result for the vector current is consistent with another renormalization method in the limit of vanishing lattice spacing.
The computation of the renormalization constant for the scalar and tensor current is underway.

\acknowledgments
We thank the members of the JLQCD collaboration for discussions and for providing the computational framework and lattice data. Numerical calculations are performed on SX-Aurora TSUBASA at High Energy Accelerator Research Organization (KEK) under its Particle, Nuclear and Astro Physics Simulation Program, as well as on Oakforest-PACS supercomputer operated by Joint Center for Advanced High Performance Computing (JCAHPC).
This work used computational resources of supercomputer Fugaku provided 
by the RIKEN Center for Computational Science through the HPCI System 
Research Project (Project ID: hp210146).
This work is supported in part by RIKEN Junior Research Associate Program, by JSPS KAKENHI Grant Numbers JP18H03710, JP21H01085, and by the Post-K and Fugaku supercomputer project through the Joint Institute for Computational Fundamental Science (JICFuS).

\bibliographystyle{unsrt}
\bibliography{lattice2021}

\end{document}